\documentclass[paper,longauth,traditabstract]{aa} 
\usepackage{graphicx}
\usepackage{txfonts}
\usepackage{natbib}
\usepackage{epstopdf}

\usepackage{longtable}
\usepackage{rotating}
\usepackage{pdflscape}
\usepackage{multirow}

\begin{document}

    \title{\emph{Herschel}\thanks{$Herschel$ is an ESA space observatory 
    with science instruments provided by 
           European-led Principal Investigator consortia and with important participation from NASA.}
            view of the Taurus B211/3 filament and striations: Evidence of filamentary growth?}

   \subtitle{}

   \author{
     P. Palmeirim\inst{1}
     \and
        Ph. Andr\'e\inst{1}
      	\and
           J. Kirk\inst{2}
	\and
	 D. Ward-Thompson\inst{3}
       \and
        	D. Arzoumanian\inst{1}
            \and       
          V. K\"onyves\inst{1,4} 	
            \and
        P. Didelon\inst{1}	
           \and
          N. Schneider\inst{1,5}		 
          \and
           M. Benedettini\inst{6}                        
           \and
           S. Bontemps\inst{5}
           \and
          J. Di Francesco\inst{7,8}
         \and
	D. Elia\inst{6}
	\and
            M. Griffin\inst{2}           
            \and	
           M. Hennemann\inst{1}      
            \and
           T. Hill\inst{1}       
          \and       
          P. G. Martin\inst{9}
          \and
            A. Men'shchikov\inst{1}
             \and
             S. Molinari\inst{6}
             \and
           F. Motte\inst{1}
             \and   
           D. Nutter\inst{2}
           \and         
        N. Peretto\inst{1}
           \and
          S. Pezzuto\inst{6}
           \and
           A. Roy\inst{9}
           \and
           K. L. J. Rygl\inst{6}
          \and
           L. Spinoglio\inst{6}
             \and
        G. White\inst{2}
            }

   \institute{Laboratoire AIM, CEA/DSM--CNRS--Universit\'e Paris Diderot, IRFU/Service d'Astrophysique, C.E.A. Saclay,
              Orme des Merisiers, 91191 Gif-sur-Yvette, France
              \email{pedro.palmeirim@cea.fr, pandre@cea.fr}
                \and
             School of Physics \& Astronomy, Cardiff University, Cardiff, UK
             \and
	   Jeremiah Horrocks Institute, University of Central Lancashire, PR1 2HE, UK
	               \and
	    Institut d'Astrophysique Spatiale, UMR 8617, CNRS/Universit\'e Paris-Sud 11,  91405 Orsay, France          
                           \and
	  Universit\'e de Bordeaux, Laboratoire d'Astrophysique de Bordeaux, CNRS/INSU, UMR 5804, BP 89, 33271 Floirac Cedex, France
            \and
             INAF - IAPS, via Fosso del Cavaliere 100, I-00133 Roma, Italy
             \and
             National Research Council of Canada, Herzberg Institute of Astrophysics,
             5071 West Saanich Road, Victoria BC Canada, V9E 2E7
            \and                      
            Department of Physics and Astronomy, University of Victoria, PO Box 355, STN CSC, Victoria BC Canada, V8W 3P6
            \and
             Canadian Institute for Theoretical Astrophysics, University of Toronto, 60 St. George Street, Toronto, ON, M5S 3H8, Canada
                }
   \date{}
   
   \abstract{
We present first results from the \textit{Herschel} Gould Belt survey for the B211/L1495 region in the Taurus molecular cloud.
Thanks to their high sensitivity and dynamic range, the \textit{Herschel} images reveal the structure of the dense, star-forming filament B211 
with unprecedented detail, along with the presence of striations perpendicular to the filament and
generally oriented along the magnetic field direction as traced by optical polarization vectors. 
Based on the column density and dust temperature maps derived from the \textit{Herschel} data, we find that 
the radial density profile of the B211 filament approaches a power-law behavior $\rho \propto r^{-2.0\pm0.4}$ at large radii 
and that the temperature profile exhibits a marked drop at small radii. 
The observed density and temperature profiles of the B211 filament are in good agreement with a theoretical model of 
a cylindrical filament undergoing gravitational contraction with a polytropic equation of state: $P \propto \rho^{\gamma}$ 
and $T \propto \rho^{\gamma-1}$, with $\gamma$=0.97$\pm$0.01$<$1 (i.e. not strictly isothermal). 
The morphology of the column density map, where some of the perpendicular striations are apparently connected to 
the B211 filament, further suggests that the material may be accreting along the striations onto the main filament. 
The typical velocities expected for the infalling material in this picture are ${\sim} 0.5$--1~km/s, which are consistent with  
the existing kinematical constraints from previous CO observations.
 }

\keywords{stars: formation -- ISM: individual objects: B211 --  ISM: clouds  -- ISM: structure -- evolution --  submillimeter: ISM  }

   \maketitle
%

\section{Introduction}

A growing body of evidence indicates that interstellar filaments play a fundamental role in the star formation process.
In particular, the results from the $Herschel$ Gould Belt survey (HGBS) confirm the omnipresence  
of parsec-scale filaments in nearby molecular clouds and suggest 
that the observed filamentary structure is directly related to the formation of prestellar cores \citep{Andre2010}.
While molecular clouds such as Taurus were already known to exhibit large-scale filamentary 
structures long before $Herschel$ \citep[cf.][]{Schneider1979,Goldsmith2008}, the $Herschel$ observations now demonstrate 
that filaments are truly ubiquitous in the cold interstellar medium (ISM) \citep[see][]{Men'shchikov2010, Molinari2010, Arzoumanian2011}.
Furthermore, the $Herschel$ results indicate that the inner width of the filaments is quasi-universal at ${\sim} 0.1$~pc \citep{Arzoumanian2011}. 
The characteristic filament width corresponds to within a factor of $\sim 2$ 
to  the sonic scale around which the transition between supersonic and subsonic turbulent motions occurs in diffuse, non-star-forming gas
and a change in the slope of the linewidth--size relation is observed 
(cf. Goodman et al. 1998; Falgarone et al. 2009; Federrath et al. 2010)\nocite{Falgarone2009}\nocite{Federrath2010}\nocite{Goodman1998}. 
This similarity suggests that the formation of filaments may result from turbulent compression of interstellar gas 
in low-velocity shocks \citep[cf.][]{Padoan2001}. 
Alternatively, the characteristic width may also be understood if interstellar filaments are formed as quasi-equilibrium structures 
in pressure balance with a typical ambient ISM pressure $P_{\rm ext} {\sim} 2$$-$5$\times$$10^4 \, \rm{K\, cm}^{-3} $ 
\citep[][Inutsuka et al. in prep.]{Fischera2012}.

The HGBS 
observations also show that the prestellar cores identified with $Herschel$ in active star-forming regions 
such as the Aquila Rift cloud \citep[cf.][]{Konyves2010} are primarily located within the densest 
filaments for which the mass per unit length exceeds the critical value \citep[e.g.,][]{Inutsuka1997}, $M_{\rm line, crit}$$=$2\, $c_s^2/G \,{\sim} \,15\, M_\odot$/pc,  
where $c_{\rm s} {\sim} 0.2$~km/s is the isothermal sound speed for $T {\sim} 10$~K. 
These $Herschel$ results support a scenario according to which core formation occurs in two main steps \citep[e.g.,][]{Andre2010,Andre2011}. 
First, large-scale magneto-hydrodynamic (MHD) turbulence gives rise to a web-like network of filaments in the ISM.
In a second step, gravity takes over and fragments the densest filaments into prestellar cores via gravitational instability.
Indirect arguments further suggest that dense, self-gravitating filaments, which are expected to undergo 
radial contraction (e.g. Inutsuka \& Miyama 1997), can maintain a constant central width of $0.1$~pc if they accrete 
additional mass from their surroundings while contracting \citep{Arzoumanian2011}.

In this letter, we present new $Herschel$ observations taken as part of the HGBS toward and around 
the B211/B213 
filament in the Taurus molecular cloud, which 
suggest that this filament is indeed gaining mass from a neighboring network of lower-density striations elongated 
parallel to the magnetic field \citep[see also][]{Goldsmith2008}. 
Owing to its close distance to the Sun \citep[$d {\sim} 140$~pc --][]{Elias1978}, the Taurus cloud has been the subject 
of numerous observational and theoretical studies. 
In particular, it has long been considered as a prototypical region and has inspired magnetically-regulated models 
of low-mass, dispersed star formation \citep[e.g.,][]{Shu1987,Nakamura2008}. 
As pointed out by \citet{Hartmann2002}, most of the young stars in Taurus are located in two or three nearly 
parallel, elongated bands, which are themselves closely associated with prominent gas filaments 
\citep[e.g.,][]{Schneider1979}. The B211/B213 filament discussed here corresponds 
to one of  these well-known star-forming filaments \citep[see also][]{Schmalzl2010, Li2012}.


\section{\textit{Herschel} observations and data reduction}

The B211/B213+L1495 area (${\sim}6^{\circ}$$ \times$$ 2.5^{\circ}$) was observed 
with \textit{Herschel}  \citep{Pilbratt2010} 
as part of the HGBS in the Taurus molecular cloud 
\citep[see ][for a presentation of early results 
obtained towards two other Taurus fields]{Kirk2012}.
Each field was mapped in two orthogonal scan directions at $60\arcsec \, {\rmÊs}^{-1}$, with both PACS \citep{Poglitsch2010} at 70~$\mu$m \& 160~$\mu$m 
and SPIRE \citep{Griffin2010} at 250~$\mu$m, 350~$\mu$m \& 500~$\mu$m, 
using the parallel-mode of \textit{Herschel}.
For B211+L1495, the North-South scan direction was split into two observations taken on 12 February 2010 
and 7 August 2010, while the East-West cross-scan direction was observed in a single run on 8 August 2010.
An additional PACS observation was taken on 20 March 2012 in the orthogonal scan direction at $60\arcsec \, {\rmÊs}^{-1}$ 
to fill a gap found in the previous PACS data.

The PACS data reduction was performed in two steps.
The raw data were first processed up to level-1 in HIPE 8.0.3384, using standard steps in the pipeline.
These level-1 data were then post-processed 
with Scanamorphos version 16 \citep{Roussel2012}, to remove glitches, thermal drifts, uncorrelated 1/$f$ noise, and produce the final maps.
The SPIRE data were reduced with HIPE 7.0.1956 using the destriper-module with a linear baseline.
The final SPIRE maps were combined using the $``\rm naive"$ map-making method, including the turn-around data. 

Zero-level offsets were added to the \textit{Herschel} maps based on a cross-correlation of the \textit{Herschel} data
with IRAS and Planck data at comparable wavelengths \citep[cf.][]{Bernard2010}.
The offset values were 2.3, 37.3, 40.2, 24.8, and 11.5 MJy/sr at 70, 160, 250, 350, and 500 $\mu$m, respectively.
A dust temperature map and a column density map ($T_{\rm dust}$ and $N_{\rm H_{2}}$ -- see online Fig.~\ref{nh2_temp_maps})
were then derived from the resulting images at the longest four  \textit{Herschel} wavelengths.  
The $N_{\rm H_{2}}$ map was reconstructed at the 18.2$''$ (0.012~pc at 140~pc) resolution of the SPIRE 250~$\mu$m data 
using the method described in Appendix A.


\onlfig{1}{
   \begin{figure*}
   \vspace{-10pt}
   \centering
   \hspace{-12mm}
  \resizebox{17cm}{!}{
 \includegraphics[angle=0]{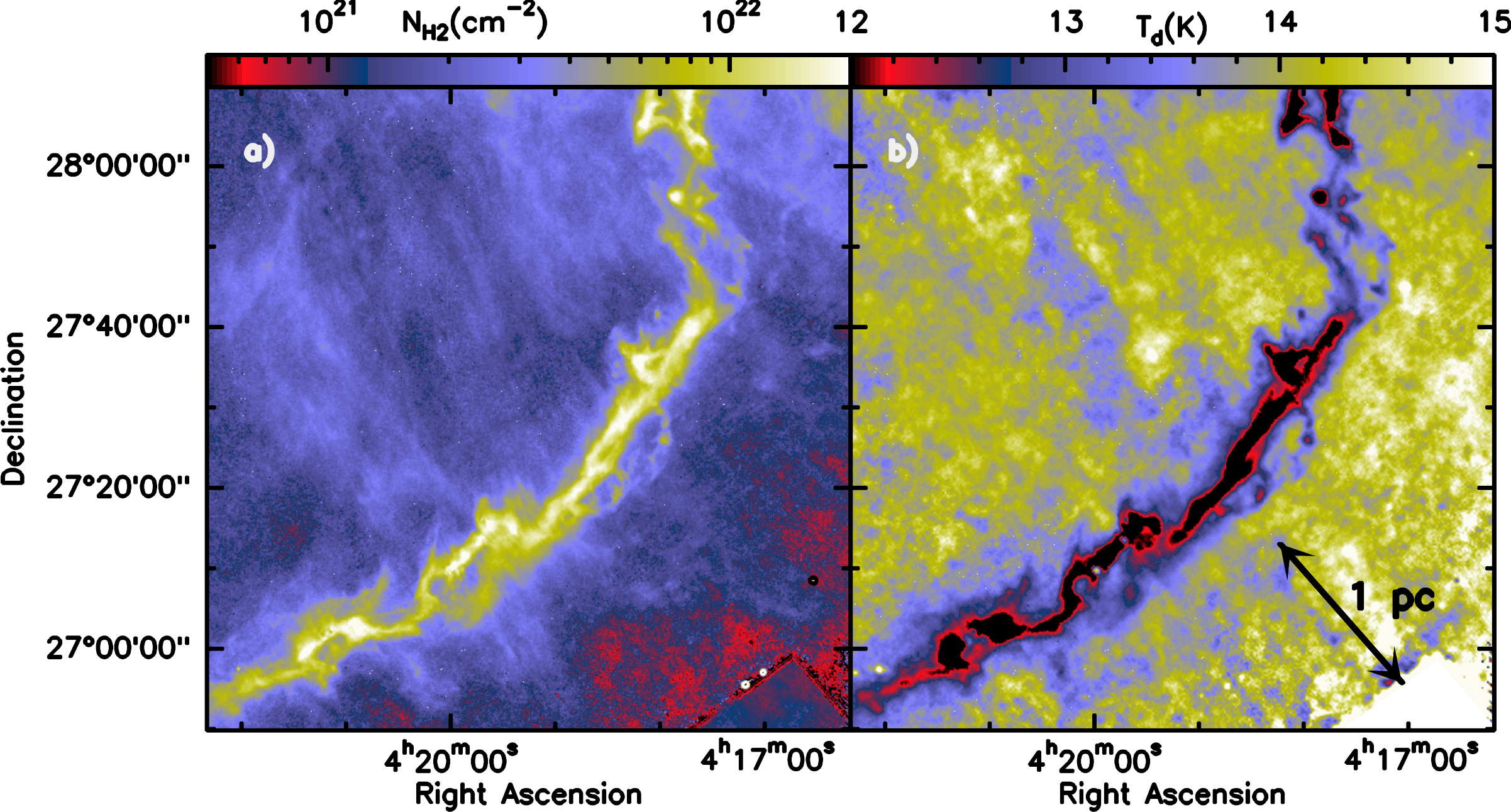}}
   \caption{ {\bf(a)} High-resolution (18.2\arcsec) column density map of the Taurus B211/B213 region (in units of $N_{\rm H_{2}}\, $cm$^{-2}$)
   derived from \textit{Herschel} data as explained in Appendix A. 
   {\bf(b)} Dust temperature map of the Taurus B211/B213 region (in K) derived at 36.3\arcsec  ~from \textit{Herschel} data.
   Comparison of the two panels shows how dust temperature and column density are anti-correlated.
}
              \label{nh2_temp_maps}
     \end{figure*}
}

  
  %
   \begin{figure}
       \begin{minipage}{1\linewidth}
     \resizebox{\hsize}{!}{
   \includegraphics[angle=0]{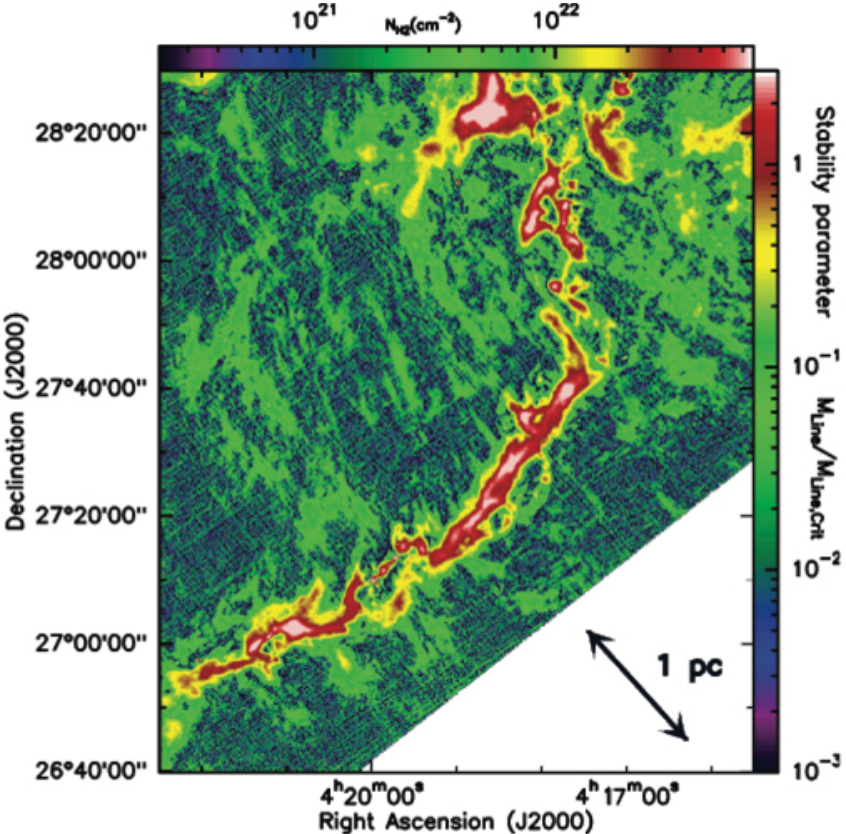}
   }
    \end{minipage}   
   \caption{High-resolution (18.2\arcsec) column density map of the Taurus B211+L1495 field derived from the \textit{Herschel} data. 
   The contrast of filamentary structures 
   has been enhanced using 
   a curvelet transform  \citep[cf.][]{Starck2003}. 
   Given the typical width ${\sim}$0.1\,pc of the filaments \citep{Arzoumanian2011},  this map is approximately equivalent 
   to a map of mass per unit length along the filaments. 
 The color bar on the right shows a line-mass scale in units of the thermal critical line mass of \citet{Inutsuka1997}, 
   which we estimate to be accurate to better than a factor of $\sim 2$ according to a detailed analysis of the radial profiles 
   of the filaments (Arzoumanian et al., in prep.). 
  Note that the main B211 filament is thermally supercritical, while the mass per unit length of the faint striations  
  is an order of magnitude below the critical value.
}
              \label{curvelet}
   \end{figure}

\section{Analysis of the filamentary structure}

In order to facilitate the visualization of individual filaments, 
we performed a ``morphological component analysis'' (MCA) decomposition 
of the \textit{Herschel} column density 
map on a basis of curvelets and wavelets \citep[e.g.,][2004]{Starck2003}\nocite{Starck2004}.
The decomposition was made on 6 scales and 100 iterations were used\footnote{The MCA software employed for this decomposition 
is publicly avaialble from the ``Inpainting routines'' link on http://irfu.cea.fr/Phocea/Vie\_des\_labos/Ast/ast\_visu.php?id\_ast=1800.}.
The map corresponding to the sum of all 6 curvelet components, shown in Fig.~\ref{curvelet}, 
provides a high-contrast view of the filaments 
after subtraction of the non-filamentary background and most of the compact cores. 
Figure.~\ref{curvelet} shows that the B211 filament is surrounded by a large number of lower-density filaments 
or striations which are oriented roughly perpendicular to the main filament.  
It is important to stress that the striations can also be seen in the original maps (cf. Figs.~ \ref{nh2_temp_maps}a and \ref{pol}a) 
and that the curvelet transform was merely used to enhance their contrast.
Some of these striations are also visible in the $^{12}$CO(1--0) and $^{13}$CO(1--0) maps of \citet{Goldsmith2008} at 45\arcsec ~resolution. 
Following \citet{Andre2010}, the column density map of Fig.~\ref{curvelet} was also converted 
to an approximate map of mass per unit length along the filaments by multiplying the local column density by 
the characteristic filament width of 0.1~pc 
(see color scale on the right of Fig.~\ref{curvelet}). 
It can be seen in Fig.~\ref{curvelet} that the mass per unit length of the main B211 filament exceeds  
the thermal value of the critical line mass, $M_{\rm line, crit}$$=$$2\, c_{\rm s}^2/G $, while the mass per unit length 
of the striations is an order of magnitude below the critical value. Assuming that the non-thermal component 
of the velocity dispersion inside the filaments is small compared to the sound speed, as suggested
by the results of millimeter line observations \citep[see][for the L1517 filament and Arzoumanian et al. 2012\nocite{Arzoumanian2012}]{Hacar2011}  
we tentatively conclude that the B211 filament is globally gravitationally unstable, while the perpendicular striations are not.

\subsection{A bimodal distribution of filament orientations}


   \begin{figure*}
   \vspace{-5pt}
   \centering
     \resizebox{\hsize}{!}{
    \includegraphics[angle=0]{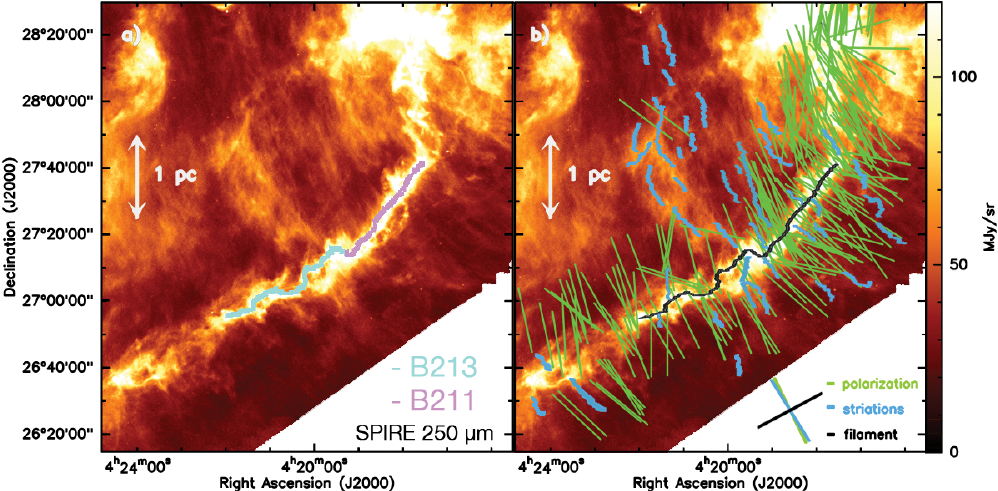}}     
   \caption{ {\bf(a)} $Herschel$/SPIRE 250 $\mu$m image of the B211/B213/L1495 region in Taurus. The light blue and purple curves 
    show the crests of the B213 and B211 segments of the whole filament discussed in this paper,  respectively.
    {\bf(b)} Display of optical and infrared polarization vectors from \citet{Heyer2008}, \citet{Heiles2000}, and \citet{Chapman2011} 
    tracing the magnetic field orientation in the B211/L1495 region, overlaid on our $Herschel$/SPIRE 250 $\mu$m image. 
   The plane-of-the-sky projection of the magnetic field appears to be oriented perpendicular to the B211/B213 filament and roughly aligned with the general direction of the striations overlaid in blue.
   The green, blue, and black segments in the lower right corner represent the average position angles of the polarization vectors, low-density  striations,  and B211 filament, respectively.  
  	}
              \label{pol}
   \end{figure*}

   \begin{figure}
       \begin{minipage}{1\linewidth}
   \centering
     \resizebox{\hsize}{!}{
    \includegraphics[angle=0]{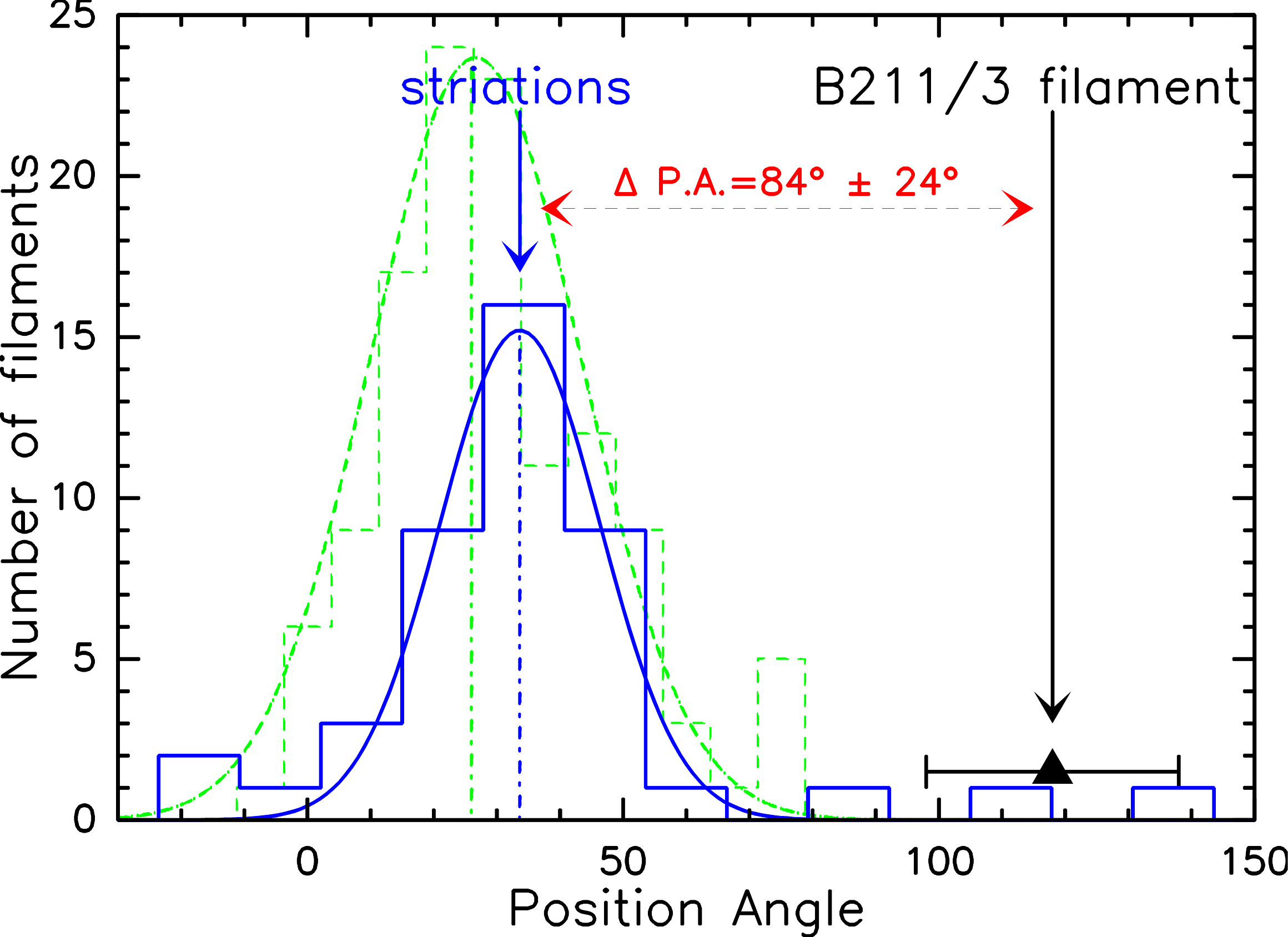}}
    \end{minipage}   
   \caption{Histogram of orientations for the low-density striations identified with DisPerSE in the B211+L1495 field (displayed in blue).
    The position-angle distribution of available optical polarization \citep[Heiles 2000]{Heyer2008} and infrared vectors \citep{Chapman2011} are also shown
    (green dashed histogram). Gaussian fits to these distributions are superimposed, indicating a peak position angle of 
    $34^{\circ}$$\pm$$13^{\circ}$ for the striations and $26^{\circ}$$\pm$$18^{\circ}$ for the B-field polarization vectors.
     The B211 filament has a mean position angle of $118^{\circ}$$\pm$$20^{\circ}$ (black triangle and horizontal error bar)
     and is thus roughly perpendicular to both the low-density striations 
     and the local direction of the magnetic field. 
 }
                 \label{pa_hist}
   \end{figure}

To analyze the distribution of filament orientations in a quantitative manner, we 
applied the DisPerSE algorithm \citep{Sousbie2010} to the original column density map (Fig.~1a), 
in order to produce a census of filaments and to trace the locations of their crests. 
DisPerSE is a general method based on principles of computational topology and it has already 
been used successfully to trace the filamentary structure in \textit{Herschel} images of star-forming 
clouds \citep[e.g.,][]{Arzoumanian2011,Hill2011,Peretto2012,Schneider2012}. 
Using DisPerSE with a relative 'persistence' threshold of $10^{21}$ c$\rm m^{-2}$ ($\sim 5\sigma $ in the map -- see Sousbie 2011 
for the formal definition of 'persistence') and an absolute column density threshold of 1--2$\, \times$$10^{21}$ c$\rm m^{-2}$, 
we could trace the crests of the B211 filament and 44 lower density
filamentary structures (see Fig.~\ref{pol}).
Due to differing background levels on either side of the B211/3 filament (see Fig.~1a), 
we adopted different column density thresholds on the north-eastern side ($2\times$$10^{21}$ c$\rm m^{-2}$)
and south-western side ($10^{21}$ c$\rm m^{-2}$). The results of DisPerSE were also visually inspected 
in both the original and the curvelet column density map, and a few doubtful features discarded.
The mean orientation or position angle of each filament was then calculated from its crest 
(see Appendix A of Peretto et al. 2012 for details). 
Figure~\ref{pa_hist} shows the resulting histogram of position angles.
In this histogram, the low-density striations are concentrated near a position angle of $34^{\circ}$$\pm$$13^{\circ}$,
which is almost orthogonal to the B211 filament (P.A.$\, =$118$^{\circ}$$\pm$$20^{\circ} $).
Interestingly, the position-angle distribution of available optical polarization vectors \citep{Heiles2000,Heyer2008}, 
which trace the local direction of the magnetic field projected onto the plane-of-sky, is centered on  P.A.$\, =26^{\circ}$$\pm$$18^{\circ}$ 
and thus very similar to the orientation distribution of the low-density striations (see Fig.~\ref{pa_hist}).
Figure~\ref{pol}b further illustrates that the low-density striations are roughly parallel 
to the B-field polarization vectors and perpendicular to the B211 filament.

\subsection{Density and temperature structure of the B211 filament}


  \begin{figure*}
      \hspace{-6mm}
  \resizebox{10.cm}{!}{
\includegraphics[angle=0]{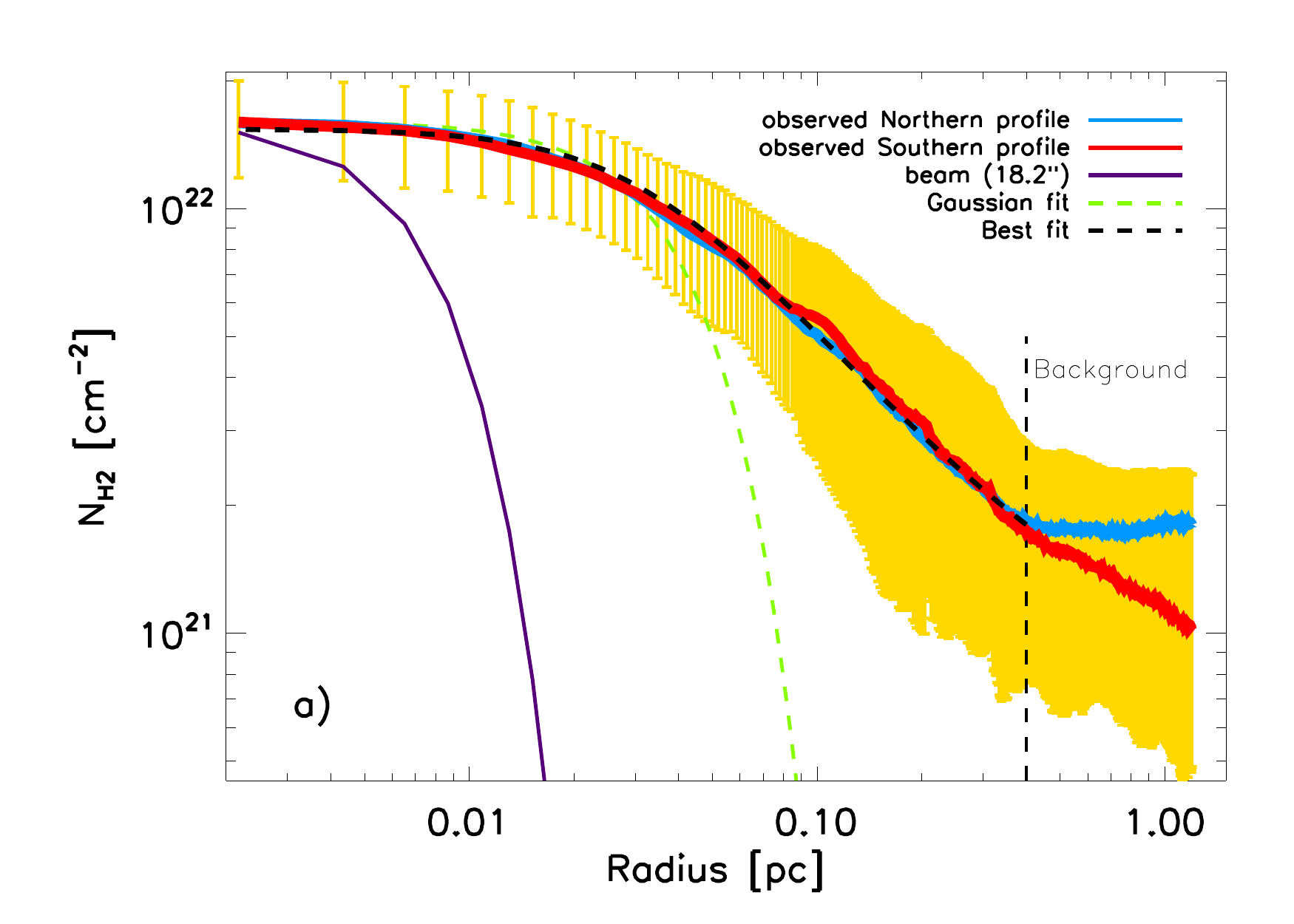}}      
      \hspace{-8mm}
  \resizebox{10.cm}{!}{
 \includegraphics[angle=0]{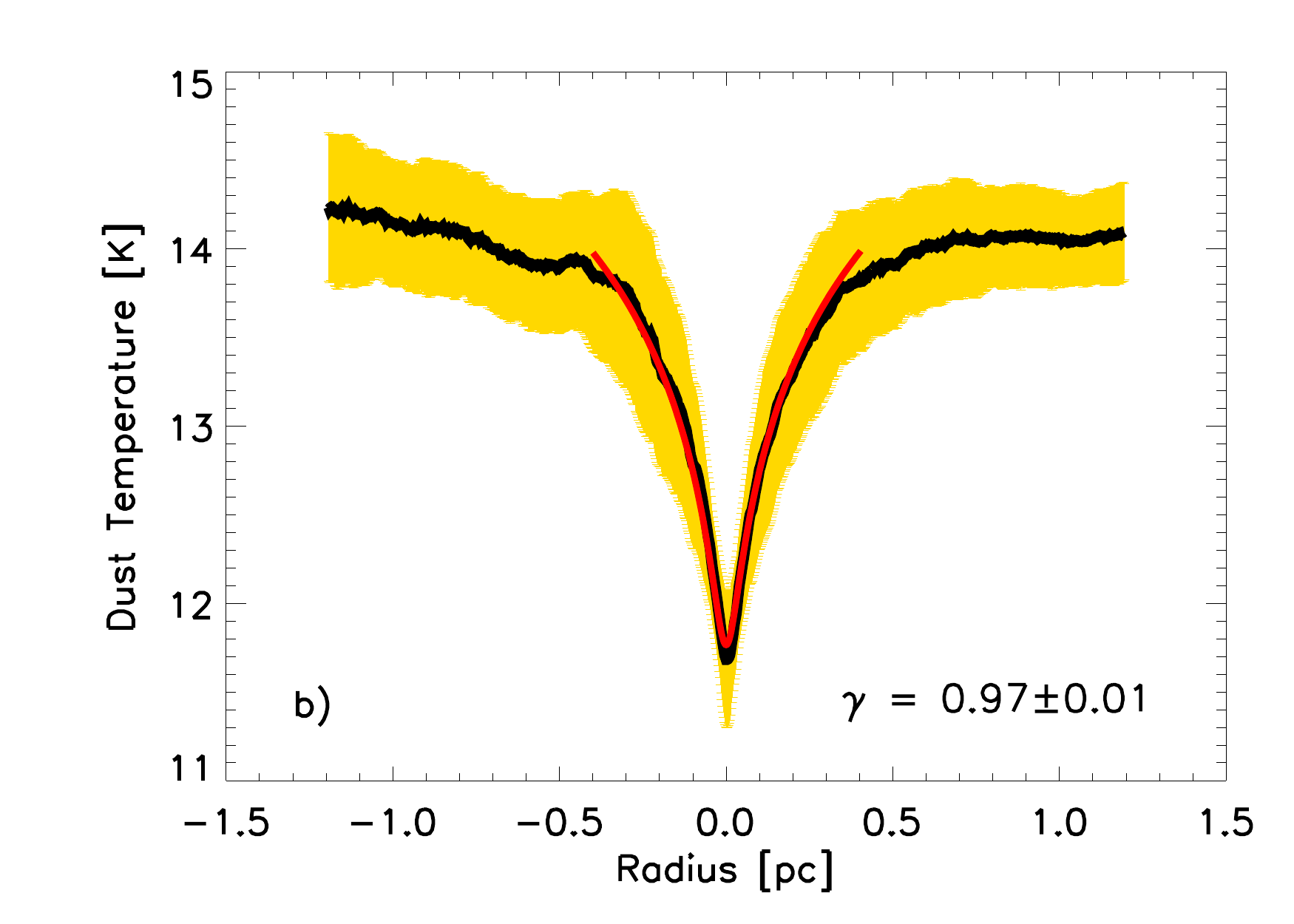}}
   \caption{{\bf(a)} Mean radial column density profile observed perpendicular to the B211 filament and displayed in log-log format, 
for both the Northern (blue curve) and the Southern part (red curve) of the filament.
The yellow area shows the ($\pm 1\sigma$) dispersion of the distribution of radial profiles along the filament.
The inner solid purple curve shows the effective 18.2\arcsec ~HPBW resolution
(0.012~pc at 140~pc) of  the column density map (online Fig.~\ref{nh2_temp_maps} - see Appendix~A for details) used to construct the profile. 
The northern and southern column density profiles are very similar up to $r$${\sim}$0.4\,pc (vertical dashed line) 
and differ significantly only for $r$$>$0.4\,pc, due to different background levels on either side of the filament.
The dashed black curve shows the best-fit Plummer model (convolved with the 18.2\arcsec ~beam) described by Eq.~(1) with $p$=2.0$\pm$0.4 and $R_{\rm \rm flat}$=0.03$\pm$0.01\,pc 
for  $r$$\leq$0.4\,pc, and including a separate linear baseline on each side representing the background for $r$$>$0.4\,pc
[see Eq.~(B1) in Appendix B for details].
The dashed curve in light green shows a Gaussian fit to the central part of the profile (mean deconvolved FWHM width ${\sim}$0.09$\pm$0.02 pc).
{\bf(b)} Mean dust temperature profile measured perpendicular to the B211 filament and displayed using a linear scale (black curve).
The solid red curve shows the best model temperature profile obtained by assuming that the filament has a density profile given 
by the Plummer model shown in {\bf(a)} and obeys a polytropic equation of state, $P \propto \rho_p^{\gamma}$, and 
thus $T(r) \propto \rho_p(r)^{(\gamma-1)}$. This best fit has  $\gamma$=0.97$\pm$0.01.
 }
              \label{fil_prof}
    \end{figure*}

Using the original 
column density and temperature maps derived from the \textit{Herschel} data (see Appendix~A),
we produced radial column density and temperature profiles for the B211 filament,
following the same procedure as \citet{Arzoumanian2011} for IC5146.
We first determined the direction of the local tangent for each pixel along the crest of the B211 filament as traced by DisPerSE.
For each pixel, we then derived one temperature profile and one column density profile 
in the direction perpendicular to the local tangent. 
Finally, by averaging all individual cuts along the crest, we obtained a mean column density 
and a mean temperature profile for the B211 filament (Fig.~\ref{fil_prof}).

To characterize the resulting column density profile we made use of an analytical model of an idealized cylindrical filament.
This model features a dense, \rm flat inner portion and approaches a power-law behaviour at large radii.
Analytically, it is described by a Plummer-like function of the form \citep[cf.][]{Nutter2008, Arzoumanian2011} :
\vspace{-0.2cm}
$$ \rho_{p}(r) = \frac{\rho_{c}}{\left[1+\left({r/R_{\rm \rm flat}}\right)^{2}\right]^{p/2}}\  \longrightarrow 
 \Sigma_{p}(r) = A_{p}\,  \frac{\rho_{\rm c}R_{\rm \rm flat}}{\left[1+\left({r/R_{\rm \rm flat}}\right)^{2}\right]^{\frac{p-1}{2}}}, \ \ (1)$$

\noindent 
where $\rho_{c}$ is the central density of the filament, $R_{\rm \rm flat}$ is the radius of the \rm flat inner region, $p$ is the power-law exponent at large
radii ($r$$>>$$R_{\rm flat}$), $A_p = \frac{1}{\cos\,i} \times \rm{B}\left(\frac{1}{2},\frac{p-1}{2}\right) $ is a finite constant factor (for $p$$>$1) that takes into account 
the filament's inclination angle to the plane of the sky (here assumed to be $i$$=$$0^{\circ}$), and $\rm{B}$ represents the Euler beta function 
\citep[cf.][]{Casali1986}. The density structure of an isothermal gas cylinder in hydrostatic equilibrium follows Eq.~(1) with $p = 4$ \citep{Ostriker1964}.

According to the best-fit model of B211 (cf. Fig.~\ref{fil_prof}a), the diameter of the \rm flat inner portion is $2\, R_{\rm flat}$$=$$0.07$$\pm$$0.02$\,pc, 
which is well resolved compared to 
the 0.012~pc (or 18.2$''$) resolution of the column density map.
The power-law regime at large radii is $\rho \propto r^{-2.0\pm0.4}$, which is significantly shallower 
than the steep $\rho \propto r^{-4}$ profile expected for unmagnetized isothermal filaments 
but would be consistent with models of isothermal equilibrium filaments threaded by helical magnetic fields (Fiege \& Pudritz 2000).
Note that our results for the density profile of the B211 filament (e.g. mean deconvolved FWHM width ${\sim}0.09$$\pm$$0.02$\,pc)
agree with the characteristic width ${\sim} 0.1$~pc  found by  \citet{Arzoumanian2011} for filaments in IC5146, Aquila, Polaris, 
and very similar to the findings of  \citet{Malinen2012} for the TMC-1 filament (also known as the Bull's tail -- Nutter et al. 2008). 
Malinen et al. derived $p$$\approx$$2.3$ and $2\, R_{\rm flat}$$\approx$$0.09$\,pc, also based on $Herschel$ data from the HGBS.
In addition, the dust temperature profile of the B211 filament shows a pronounced temperature drop 
toward the center  (Fig.~\ref{fil_prof}b), which suggests that the 
gas is not strictly isothermal$^{2}$. 

A reasonably good model for the 
structure of the B211 filament is obtained by considering 
similarity solutions for the collapse of an infinite cylinder obeying a polytropic (non-isothermal) equation of state 
of the form $P \propto \rho^\gamma$ with $\gamma$$\la$$1$. \citet{Kawachi1998} have shown that the outer 
density profile of such a collapsing cylinder approaches the power law $\rho \propto r^{-\frac{2}{2-\gamma}}$.
For $\gamma$ values close to unity, the model column density profile thus approaches $\rho \propto r^{-2}$  at large radii, 
which is consistent with the observed profile of the B211 filament.  
The best Plummer model derived above for the density profile [$\rho_p(r)$ -- Fig.~\ref{fil_prof}a] 
can be used to estimate the $\gamma$ value which leads to the best fit to 
the observed dust temperature profile\footnote{We make the approximation that $T_{gas}(r) \approx T_{dust}(r)$, which should be correct in the inner part of the B211 filament at least, since the 
gas and dust temperatures are expected to be well coupled in high-density ($\geq$$3$$\times$10$^4$ cm$^{-3}$) regions \citep[see][]{Galli2002} 
and the central density of the filament is estimated to be $n_c$$\approx$4.5$\times$10$^4$ cm$^{-3}$.} 
under the polytropic assumption [$T(r) \propto \rho_p(r)^{(\gamma-1)}$].
The resulting model fit, overlaid in red in Fig.~\ref{fil_prof}b, has $\gamma$$=$$0.97$$\pm$$0.01$$<$$1$, corresponding 
to  $\rho \propto r^{-1.96 \pm 0.02} $ at large radii for a self-similar, not strictly isothermal collapsing cylinder.


\section{Discussion: Contraction and accretion in B211?} 
 
The results presented in this paper reveal the density and temperature structure of the Taurus B211 filament with unprecedented detail.
The shape of the column density profile derived for the B211 filament, with a well-defined power-law regime at large radii 
(see Fig.~\ref{fil_prof}a), and the large column density contrast over the surrounding background 
(a factor ${\sim} 10$$-$$20$, implying a density contrast $ {\sim} 100$$-$$400$) 
strongly suggest that the main filament has undergone gravitational contraction. This is also consistent with the supercritical 
mass per unit length measured for the B211 filament ($M_{\rm line}$$\approx$$54\, M_\odot $/pc), which 
suggests that the filament is unstable to both global radial contraction and fragmentation into cores 
\citep[e.g.,][]{Inutsuka1997,Pon2011}.
Observations confirm that the B211 filament has indeed fragmented, leading to the formation of several prestellar cores \citep[e.g.,][]{Onishi2002} 
and protostars \citep[e.g.,][]{Motte2001,Rebull2010} along its length.

The orientation alignment of the striations with optical polarization vectors suggests that 
the magnetic field plays an important role in shaping the morphology of the filamentary structure in this part of Taurus.
Earlier studies, using similar polarization observations of background stars, already pointed out that the structure of the Taurus cloud was strongly correlated with 
the morphology of the ambient magnetic field (e.g. Heyer et al. 2008; Chapman et al. 2011 and references therein). 
Using the Chandrasekhar--Fermi method, \citet{Chapman2011} estimated a magnetic field strength of ${\sim} 25\, \mu$G 
in the B211 area and concluded that the region corresponding to the striations seen here in, e.g., Fig.~\ref{curvelet} and Fig.~\ref{pol} 
was magnetically subcritical. 

Theoretical arguments \citep[e.g.,][]{Nagai1998} predict that, in the presence of a ``strong'' magnetic field, low-density, thermally subcritical filaments
such as the striations observed in Taurus should be preferentially oriented parallel to the field lines, while high-density, self-gravitating filaments 
should be preferentially oriented perpendicular to the field lines. 
This difference arises because low-density structures 
which are not held by gravity 
have a tendency to expand and disperse, while self-gravitating 
structures have a tendency to contract. 
In the presence of a magnetic field, 
motions of slightly ionized gas do not encounter any resistance along the field lines 
but  encounter significant resistance perpendicular to the field lines.  
Consequently, an initial perturbation in a low-density part of the cloud will tend to expand along the field lines 
and form an elongated structure or a subcritical ``filament'' parallel to the field. 
Conversely, a self-gravitating structure will tend to contract along the field lines, forming a condensed, self-gravitating sheet \citep[cf.][]{Nakamura2008}
which can itself fragment into several supercritical filaments oriented perpendicular to the field \citep[e.g.][]{Nagai1998}.
These simple arguments may explain the distribution of filament orientations with two orthogonal groups found in Sect.~3.1 (see Fig.~4). 
Other regions imaged with \textit{Herschel} where a similar distribution of filament orientations is observed and a similar 
mechanism may be at work include the Pipe nebula \citep{Peretto2012}, the Musca cloud (Cox et al. in prep.) 
and the DR21 ridge in Cygnus X \citep{Schneider2010,Hennemann2012}.

\begin{figure} 
      \begin{minipage}{1\linewidth}
   \centering
     \resizebox{\hsize}{!}{   
   \includegraphics[angle=0]{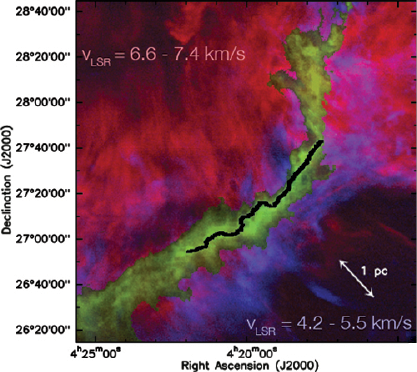}}
       \end{minipage}    
   \caption{CO emission observed toward and around the B211 filament (Goldsmith et al. 2008). 
Redshifted $^{12}$CO(1--0) emission integrated from $V_{LSR}$$=$$6.6$\,km/s to $V_{LSR}$$=$$7.4$\,km/s is displayed in red and mostly seen to 
the north-east of the B211 filament. 
Blueshifted $^{12}$CO(1--0) emission integrated from $V_{LSR}$$=$$4.2$\,km/s to $V_{LSR}$$=$$5.5$\,km/s is displayed in blue and mostly seen to 
the south-west of the filament. 
The main body of the B211 filament, displayed in green, corresponds to the $^{13}$CO(1--0) emission detected between 
$V_{LSR}$$=$$5.6$\,km/s to $V_{LSR}$$=$$6.4$\,km/s.
Both $^{12}$CO(1--0) and $^{13}$CO(1--0) maps have a spectral resolution of ${\sim}0.2$ km/s.
}
              \label{rsb}    
     \end{figure}

The morphology of the region with a number of low-density striations parallel to the magnetic field lines, some of them approaching 
the B211 filament from the side and apparently connected to it (see Fig.~\ref{curvelet}), is also suggestive of mass accretion 
along the field lines into the main filament. 
To test this hypothesis, we assume cylindrical geometry and use the observed mass per unit length $M_{line}$  
to estimate the gravitational acceleration $g(R)$$=$$2\,G M_{line}(r$$<$$R)/R $ of a piece of gas in free-fall toward the B211 filament, 
where $R$ represents radius.
The free-fall velocity v$_{\rm ff} $ of gas initially at rest at a cylindrical radius $R_{\rm init} {\sim} 2$~pc (corresponding to the most distant 
striations) is estimated to reach v$_{\rm ff}$$=$$2\,\left[G M_{\rm line}\, {\rm ln}(R_{\rm init}/R)\right]^{1/2}$$\approx$$1.1$\,km/s when 
the material reaches the outer radius $R {\sim} 0.4$~pc of the B211 filament. 
This free-fall estimate is an upper limit since it neglects any form of support against gravity. 
A more conservative estimate can be obtained by considering the similarity solution found by \citet{Kawachi1998} for the gravitational collapse 
of a cylindrical filament supported by a polytropic pressure gradient with $\gamma$$\la$$1$. 
In this model, the radial infall velocity in the outer parts of the collapsing filament is expected to be v$_{\rm inf} {\sim} \,$0.6--1\,km/s 
when $\gamma =\, $0.9--0.999 and the gas temperature is ${\sim} 10$\,K  (see Figs.~4 and 6 of Kawachi \& Hanawa). 
The above two velocity estimates can be compared with the kinematical constraints provided by the $^{12}$CO(1--0) observations of \citet{Goldsmith2008}. 
It can be seen in online Fig.~\ref{rsb} that there is an average velocity difference of ${\sim}1$ km/s 
between the red-shifted CO emission observed at $V_{LSR}$${\sim}$$7$\, km/s  to the north-east 
and the B211 filament which has $V_{LSR}$${\sim}$$6$\, km/s. Likewise, there is an average difference of ${\sim}1$ km/s 
between the blue-shifted CO  emission observed at $V_{LSR}$${\sim}$$5$\, km/s  to the south-west and the B211 filament.
Although projection effects may somewhat increase the magnitude of the intrinsic velocity difference,
we conclude that there is good qualitative agreement between the estimated inflow velocity in the striations and the $^{12}$CO observational constraints.
Considering these velocities, 
the current mass accretion rate onto the 4-pc-long filament (total mass of ${\sim}220M_\odot$) is estimated to be on the order of \.{M}$_{\rm line}{=\rho_{\rm p}(R) \times \rm v_{inf} \times {2\pi}R\approx}$\,27-50${\,M_\odot}$/pc/Myr, where $\rho_{\rm p}(R)$ corresponds to the density of the best-fit Plummer model at the filament outer radius R=0.4pc.
This would mean that it would take ${\sim}$1-2~Myr for the central filament to form at the current accretion rate and
${\sim}$0.8-1.5~Myr for the total mass of the striations (${\sim}150M_\odot$) to be accreted. 
The available observational evidence therefore lends some credence to the view that the B211 filament is radially contracting toward its long axis, 
while at the same time accreting additional ambient material through the striations.

\begin{acknowledgements}
Pedro Palmeirim is funded by the Funda\c{c}\~ao para a Ci\^encia e a Tecnologia (Portugal). 
We are grateful to Paul Goldsmith for making the FCRAO CO(1--0) data of the B211/L1495 region available 
to us.
We thank Shu-ichiro Inutsuka and Fumitaka Nakamura for insightful discussions about filaments.
D.E. and K.L.J.R. are funded by an ASI fellowship under contract
number I/005/11/0.
SPIRE has been developed by a consortium of institutes led by
Cardiff Univ. (UK) and including Univ. Lethbridge (Canada);
NAOC (China); CEA, LAM (France); IFSI, Univ. Padua (Italy);
IAC (Spain); Stockholm Observatory (Sweden); Imperial College
London, RAL, UCL-MSSL, UKATC, Univ. Sussex (UK); Caltech, JPL,
NHSC, Univ. Colorado (USA). This development has been supported
by national funding agencies: CSA (Canada); NAOC (China); CEA,
CNES, CNRS (France); ASI (Italy); MCINN (Spain); SNSB (Sweden);
STFC (UK); and NASA (USA).  
PACS has been developed by a consortium of institutes led by MPE
(Germany) and including UVIE (Austria); KUL, CSL, IMEC (Belgium); CEA,
OAMP (France); MPIA (Germany); IFSI, OAP/AOT, OAA/CAISMI, LENS, SISSA
(Italy); IAC (Spain). This development has been supported by the funding
agencies BMVIT (Austria), ESA-PRODEX (Belgium), CEA/CNES (France),
DLR (Germany), ASI (Italy), and CICT/MCT (Spain).

\end{acknowledgements}

\bibliography{AA}

\Online
\begin{appendix} 

\section{Derivation of a high-resolution (18.2\arcsec) column density map}

\label{app_cd}

The procedure employed here to construct a column density map at  the 18.2\arcsec resolution of the SPIRE 250 $\mu$m data 
for the B211+L1495 region is consistent with, but represents an improvement over, the method used in earlier HGBS papers 
to derive column density maps at the 36.3\arcsec ~resolution of SPIRE 500 $\mu$m observations.
Following the spirit of a multi-scale decomposition of the data (cf. Starck et al. 2004), 
the gas surface density distribution of the region, smoothed to the resolution of the SPIRE 250~$\mu$m observations, may be expressed as a sum of 
three terms:  $$ \Sigma_{250} =  \Sigma_{500} +  \left(\Sigma_{350} - \Sigma_{500}\right) +  \left(\Sigma_{250} - \Sigma_{350}\right). \ \ (A.1)$$ 

\noindent
In the above equation, $\Sigma_{500}$,  $\Sigma_{350}$, and $\Sigma_{250}$ represent smoothed versions of the 
intrinsic gas surface density distribution $\Sigma$ after convolution with the SPIRE beam at 500 $\mu$m, 350 $\mu$m, and 250 $\mu$m, 
respectively, i.e.:  $\Sigma_{500} = \Sigma * B_{500}$, $\Sigma_{350} = \Sigma * B_{350} $, and $\Sigma_{250} = \Sigma * B_{250}$.

The first term of Eq.~(A.1)  is simply the surface density distribution smoothed to the resolution of the SPIRE 500 $\mu$m data. 
An estimate,  $\bar{\Sigma}_{500} $, of this term can be derived from the $Herschel $ data using the same 
procedure as in earlier HGBS papers (e.g. K\"onyves et al. 2010). Briefly,  the  {\it Herschel} images including the zero-level offsets 
estimated from IRAS and Planck (cf. Bernard et al. 2010) are first smoothed 
to the 500~$\mu$m resolution (36.3\arcsec) and reprojected onto the same grid. 
An optically thin greybody function of the form
$I_{\nu}$ = $B_{\nu}(T_{\rm d})\, \kappa_{\nu}\, \Sigma$, where $I_{\nu}$ is the observed surface brightness 
at frequency $\nu $, and $\kappa_{\nu} $ is the dust opacity per unit (dust$+$gas) mass, is then fitted to the spectral energy distributions (SEDs) 
observed with $Herschel$ between 160 ~$\mu$m and 500~$\mu$m, on a pixel-by-pixel basis (four SED data points per pixel). This makes it possible 
to estimate the best-fit value $\bar{\Sigma}_{500}(x,y) $ and $T_{\rm d, 500}(x,y) $ at each pixel position $(x,y)$. 
The following dust opacity law, very similar to that advocated by \citet{Hildebrand1983} at submillimeter wavelengths, is assumed:  
$\kappa_{\nu} = 0.1 \times (\nu/1000~{\rm GHz})^{\beta} =  0.1 \times (300~{\rm \mu m}/\lambda)^{\beta}$~cm$^2$/g, with $\beta = 2$. 

The second term of Eq.~(A.1) may be written as $\Sigma_{350} - \Sigma_{350}*G_{500\_350} $, 
where $G_{500\_350} $ is a circular Gaussian with full width at half maximum (FWHM) $\sqrt{36.3^2- 24.9^2} \approx 26.4\arcsec $. 
(To first  order, the SPIRE beam at 500 $\mu$m is a smoothed version of the SPIRE beam at 350 $\mu$m, i.e., 
$B_{500} = B_{350}*G_{500\_350} $.)
The second term of Eq.~(A.1) may thus be viewed as a term adding information on spatial scales accessible 
to SPIRE observations at  $350 \mu$m, but not to SPIRE observations at  $500 \mu$m. 
In practice, one can derive and estimate $\bar{\Sigma}_{350} $ of $\Sigma_{350} $ in a manner similar to $\bar{\Sigma}_{500} $, 
through pixel-by-pixel SED fitting to three $Herschel$ data points between 160 ~$\mu$m and 350~$\mu$m (i.e., ignoring the lower resolution 500~$\mu$m data point). 
An estimate of the second term of Eq.~(A.1) can then be obtained by subtracting a smoothed version of $\bar{\Sigma}_{350} $
(i.e., $\bar{\Sigma}_{350}*G_{500\_350} $) to $\bar{\Sigma}_{350} $ itself, i.e., by removing low spatial frequency information from $\bar{\Sigma}_{350} $.

Likewise, the third term of Eq.~(A.1) may be written as $\Sigma_{250} - \Sigma_{250}*G_{350\_250} $, where 
$G_{350\_250} $ is a circular Gaussian with FWHM $\sqrt{24.9^2-18.2^2} \approx 17.0\arcsec $, and may be 
understood as a term adding information on spatial scales only accessible to $Herschel$ observations at wavelengths $\leq 250 \mu$m. 
In order to derive an estimate $\bar{\Sigma}_{250} $ of $\Sigma_{250} $ on the right-hand side of Eq.~(A.1), 
we first smooth the PACS 160 $\mu$m map to the 18.2\arcsec ~resolution of the SPIRE 250 $\mu$m 
map and then derive a color temperature map between 160~$\mu$m and 250$\mu$m from the observed $I_{\rm 250 \mu m }(x,y)/I_{\rm 160 \mu m }(x,y)$ 
intensity ratio at each pixel $(x,y)$. The SPIRE 250 $\mu$m map is converted into a gas surface density map ($\bar{\Sigma}_{250} $), assuming 
optically thin dust emission at the temperature given by the color temperature map and a dust opacity at 250 $\mu$m 
$\kappa_{\rm 250 \mu m } = 0.1 \times (300/250)^2$~cm$^2$/g. 
An estimate of the third term of Eq.~(A.1) can then be obtained by subtracting a smoothed version of $\bar{\Sigma}_{250} $ 
(i.e., $\bar{\Sigma}_{250} *G_{350\_250} $) to $\bar{\Sigma}_{250} $ itself, i.e., by removing low spatial frequency information from $\bar{\Sigma}_{250} $.

Our final estimate $ \tilde{\Sigma}_{250}$ of the gas surface density distribution at 18.2\arcsec ~resolution is produced by adding up 
the above estimates of the three terms on the right-hand side of Eq.~(A.1): 
$$ \tilde{\Sigma}_{250} =  \bar{\Sigma}_{500} +  \left(\bar{\Sigma}_{350} - \bar{\Sigma}_{350}*G_{500\_350}\right) 
+  \left(\bar{\Sigma}_{250}  - \bar{\Sigma}_{250}*G_{350\_250} \right). \ \ (A.2)$$

The resulting 18.2\arcsec-resolution column density map $\tilde{N}_{\rm H_{2}}$
for the B211+L1495 region is displayed in online Fig.~\ref{nh2_temp_maps}a 
in units of mean molecules per $\rm{cm}^{2}$, where $ \tilde{\Sigma}_{250}  = \mu \, {\rm m_H} \, \tilde{N}_{\rm H_{2}}$ and $\mu = 2.33$ is the mean molecular weight.
Although this high-resolution map is somewhat noisier than its 36.3\arcsec -resolution counterpart (corresponding to $\bar{\Sigma}_{500} $) 
and has lower signal-to-noise ratio than the SPIRE 250 $\mu$m image (cf. Fig.~3), 
due to additional noise coming from the second and third terms of Eq.~(A.1), the quality and dynamic range of the $Herschel$ data 
are such that the result provides a very useful estimate of the column density distribution in B211+L1495 with a factor of 2 better 
resolution than standard column density maps derived so far from $Herschel$ observations. 

Because the higher-resolution terms in Eq.~(A.1) are derived using fewer and fewer SED data points to estimate the effective 
dust temperature $T_d$ for each line of sight,  the high-resolution column density map is also somewhat less reliable 
than the standard 36.3\arcsec -resolution column density map (corresponding to $\bar{\Sigma}_{500} $). 
To evaluate the reliability of the $\bar{\Sigma}_{350} $ and $\bar{\Sigma}_{250} $ maps entering the calculation of $ \tilde{\Sigma}_{250}$ 
[cf. Eq.~(A.2)] and derived using only three and two SED data points per position, respectively, we made the following tests 
in the case of the Taurus B211/B213+L1495 data. 
First, from the $Herschel$ 160~$\mu$m to 350~$\mu$m images smoothed to the 500~$\mu$m resolution, 
we derived a dust temperature map $T_{\rm d, 350}^{\rm 500}  $ and a gas surface density map $\bar{\Sigma}_{350}^{500} $ 
using the three $Herschel$ data points between 160 $\mu$m and 350~$\mu$m and compared these maps  
to the maps $T_{\rm d, 500}$ and $\bar{\Sigma}_{500} $ derived at the same resolution using four SED data points per position.
In the case of the Taurus data, the $T_{\rm d, 350}^{\rm 500} $ map agrees 
with the $T_{\rm d, 500}$ map to better than 0.15~K on average (and better than 0.8~K everywhere) 
and $\bar{\Sigma}_{350}^{500} $ agrees with $\bar{\Sigma}_{500} $ to better than 4\% on average (and better than 25\% everywhere).
Likewise, using the $Herschel$ 160 ~$\mu$m and 250~$\mu$m images smoothed to the 500~$\mu$m resolution, 
we derived a color temperature map $T_{\rm d, 250}^{\rm 500}  $ and a gas surface density map $\bar{\Sigma}_{250}^{500} $ 
in the same way as we calculated $\bar{\Sigma}_{250} $ above and then compared these maps 
to the $T_{\rm d, 500}$ and $\bar{\Sigma}_{500} $ maps. 
The $T_{\rm d, 250}^{\rm 500} $ map agrees with the $T_{\rm d, 500}$ map to better than 0.15~K on average (and better than 1.2~K everywhere) 
and $\bar{\Sigma}_{250}^{500} $ agrees with $\bar{\Sigma}_{500} $ to better than 3\% on average (and better than 30\% everywhere)\footnote{In pathological 
situations, such as when warm foreground dust emission from a photon-dominated region is present in front of colder structures, 
the difference maps $T_{\rm d, 350}^{\rm 500} - T_{\rm d, 500}$ and $T_{\rm d, 250}^{\rm 500} - T_{\rm d, 500}$ could potentially 
be used to improve the estimates of the second and third terms of  Eq.~(A.1). Since these difference maps remain small in the case of Taurus, we refrained from 
using them here and did not apply any correction to $\bar{\Sigma}_{350} $ and $\bar{\Sigma}_{250} $.}.
Finally, to test the robustness of  the 18.2\arcsec-resolution column density map $ \tilde{\Sigma}_{250}$, we smoothed it to the 36.3\arcsec ~resolution 
of the standard column density map and inspected the ratio map between the two, which has a mean value of 1.00 and 
a standard deviation of 0.04. Within the region covered by both PACS and SPIRE,  the smoothed version of $ \tilde{\Sigma}_{250}$ 
agrees with $\bar{\Sigma}_{500} $  to better than $10\%$.

%

\section{Details of the procedure used to fit the column density profile of the B211/B213 filament}

The fitting analysis of the observed column density profile (Sect. 3.2 and Fig.~5) was 
performed using the non-linear least-squares fitting IDL procedure MPFIT (Markwardt, C. B. 2008 - http://purl.com/net/mpfit).
%
In addition to the  Plummer-like cylindrical model filament corresponding to Eq.~(1) with three free parameters, 
the background gas was represented by two separate linear baselines 
on either side of the filament. 
Each side was fitted independently with the following model function (convolved with the beam):
$$ \Sigma_{p} (r)/\mu\, m_H = \,  \frac{N^0_{\rm H_{2}}}{\left[1+\left({r/R_{\rm \rm flat}}\right)^{2}\right]^{\frac{p-1}{2}}} + Bkg[1]r+Bkg[0], \ \ (\rm B.1)$$

\noindent
where $N^0_{\rm H_{2}}  (=A_{\rm p}  \rho_{\rm c}  R_{\rm \rm flat}/\mu\, m_H$),  R$_{\rm \rm flat}$, $p$, Bkg[1], and Bkg[0] were treated as five free parameters.
Bkg[1] and Bkg[0] are two parameters which describe the local background cloud gas. 
The results of these model fits are given in Table~\ref{pfit}.1 for both the north-eastern and the south-western side of the B211/B213 filament considered as a whole (see Fig.~\ref{fil_prof}a), 
as well as for the B211 and the B213 segment of the whole filament (see online Fig.~\ref{B211_B213}).
Due to differing background levels on either side of the filament, fitting the two sides of the filament separately 
gives better results than a global fit to both sides simultaneously.


\begin{table*}  
\begin{minipage}{1\linewidth}     
\centering
 \caption{Parameters of the Plummer-like model fits to the column density profiles of the B211/B213 filament and segments}
\begin{tabular}{c|c|c|c|c|c|c}   
\hline\hline   
Mean $N_{\rm H_{2}} $   & $N^0_{\rm H_{2}} $$^{\bf(a)}$   &   $R_{\rm flat}$   &   $p$  & Bkg[1]$^{\bf(b)}$& Bkg[0] & $\chi$$^{2}$ \\
radial profiles & [$10^{21}$ c$\rm m^{-2}$]  &  [pc] & & [$10^{21}$ c$\rm m^{-3}$]     &[$10^{21}$ c$\rm m^{-2}$]  & \\
\hline  
      north-eastern side of  B211/B213     & $14.8 \pm 1.9 $ 	   &  $0.031 \pm 0.012$   &  $ 2.03 \pm 0.34  $   & $0.86 \pm 0.45$ & $0.52 \pm 0.80$ &  1.87     \\ 
      south-western side of B211/B213   & $14.4 \pm 1.4 $ 	   &  $0.032 \pm 0.014$   &  $ 2.00 \pm 0.09  $   &$0.0$ & $0.67 \pm 0.17 $ &  2.12     \\   
\hline           
       north-eastern side of B211 segment      & $14.9 \pm 1.2 $ 	   &  $0.050 \pm 0.014$   &  $ 2.51 \pm 0.45  $    &  $0.73 \pm 0.38$ & $0.95 \pm 0.52 $ & 4.37   \\ 
       south-western side of B211 segment      & $14.0 \pm 1.0 $ 	   &  $0.057 \pm 0.015$   &  $ 2.64 \pm 0.50  $    &  $0.23 \pm 0.36$ & $1.31 \pm 0.48$ &  3.89 \\ 
\hline           
       north-eastern side of B213 segment       & $13.0 \pm 1.8 $ 	   &  $0.025 \pm 0.008$   &  $ 1.75 \pm 0.07  $    &  $0.95 \pm 0.11$ & $0.0$ & 1.41   \\ 
      south-western side of  B213 segment     & $12.7 \pm 2.0 $ 	   &  $0.020 \pm 0.007$   &  $ 1.64 \pm 0.05  $    &  $0.03 \pm 0.12$ & $ 0.0 $ &  6.00 \\     
                           \hline  
                  \end{tabular}
\vspace*{-0.45ex}
 \begin{list}{}{}
 \item[]{{\bf Notes:}  (a) Mean central column density measured along the crest of the filament (after background subtraction).
(b) On the south-western side of the B211/B213 filament, the background is well described by a constant value, i.e., Bkg[1]~$= 0$.
The fitting analysis was performed on the mean column density profiles measured on the north-eastern and south-western sides of the global (B211/B213) filament 
and the two (B211 and B213) segments (see online Fig.~\ref{B211_B213}).
}
 \end{list}      
 \end{minipage}
 \label{pfit} 
\end{table*}

\onlfig{7}{
   \begin{figure*}
   \centering
     \resizebox{16cm}{!}{   
   \includegraphics[angle=0]{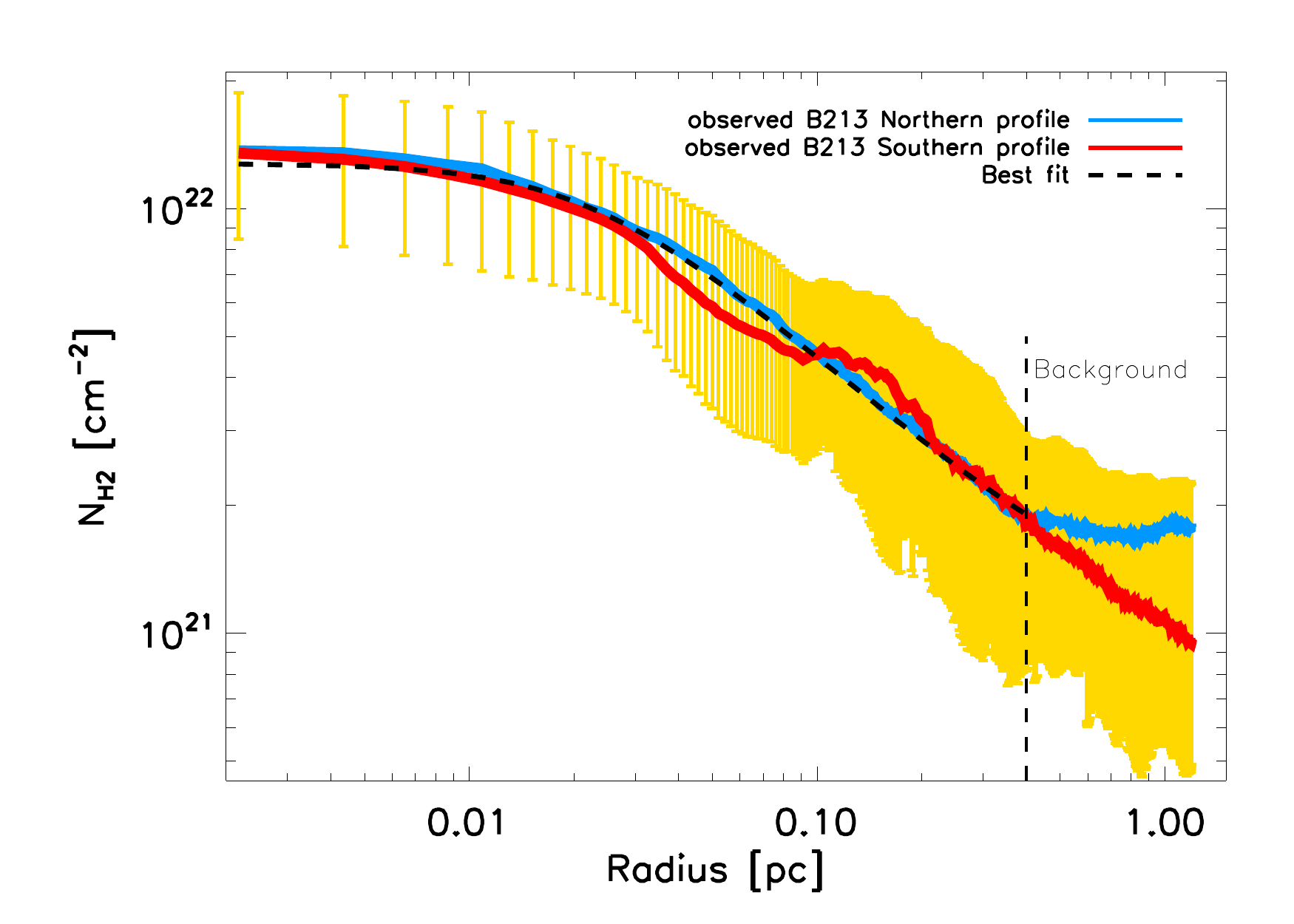}   
   \hspace{-2mm}
  \includegraphics[angle=0]{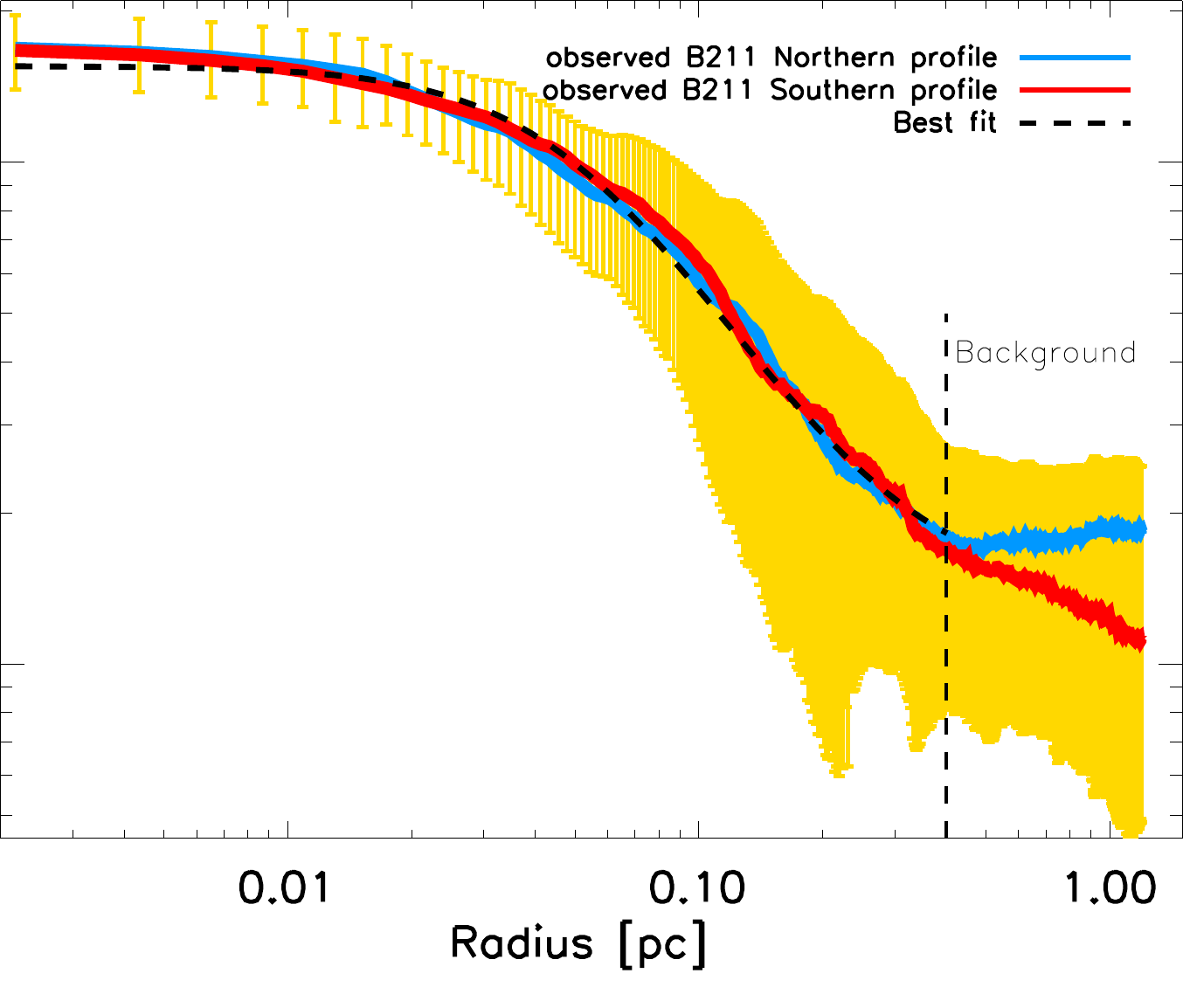} } 
   \caption{Mean radial column density profiles of the B213 (left) and B211 (right) segments of the filament for both the north-eastern (in blue) and south-western (in red) sides. 
   In both panels, the black dashed curve represents the mean of the best-fit Plummer models to the north-eastern and south-western profiles, 
   truncated at $r = 0.4 $pc where the backgrounds start to diverge significantly between the two sides.
   The crests  defining the B213 and B211 segments considered here can be seen in Fig.~\ref{pol}a. 
   The B213 segment has a slightly shallower profile ($p \approx 1.7$) and a smaller flat inner radius ($R_{\rm flat} \approx 0.025$~pc) than 
   the  B211 segment ($p \approx 2.6$ and $R_{\rm flat} \approx 0.06$~pc). The detailed parameters of the model fits are given in Table~\ref{pfit}.1. 
   }
              \label{B211_B213}
     \end{figure*}
}

\end{appendix}

\end{document}